# Feasibility of Heart Sound Analysis in Individuals Supported with Left Ventricular Assist Devices

Xinlin J. Chen[1], Emma T. LaPorte[2], Leslie M. Collins[3], Priyesh Patel[4],
Ravi Karra[5], Boyla O. Mainsah[3]

[1] Department of Electrical and Computer Engineering, Duke University, Durham, NC, USA. Corresponding author. xinlin.chen@duke.edu.
[2] SHYFT Analytics, Boston, MA, USA
[3] Department of Electrical and Computer Engineering, Duke University, Durham, NC, USA
[4] Sanger Heart and Vascular Institute, Atrium Health, Charlotte, NC, USA
[5] Division of Cardiology, Department of Medicine, Duke University Medical Center, Durham, NC, USA

## ABSTRACT

Left ventricular assist devices (LVADs) are surgically implanted mechanical pumps that improve survival rates for individuals with advanced heart failure. While life-saving, LVAD therapy is also associated with high morbidity, which can be partially attributed to the difficulties in identifying an LVAD complication before an adverse event occurs. Methods that are currently used to monitor for complications in LVAD-supported individuals require frequent clinical assessments at specialized LVAD centers. Remote analysis of digitally recorded precordial sounds has the potential to provide an inexpensive point-of-care diagnostic tool to assess both device function and the degree of cardiac support in LVAD recipients, facilitating real-time, remote monitoring for early detection of complications. To our knowledge, prior studies of precordial sounds in LVAD-supported individuals have analyzed LVAD noise rather than intrinsic heart sounds, due to a focus on detecting pump complications, and perhaps the obscuring of heart sounds by LVAD noise. In this letter, we describe an adaptive filtering method to remove sounds generated by the LVAD, making it possible to automatically isolate and analyze underlying heart sounds. We present preliminary results describing acoustic signatures of heart sounds extracted from *in vivo* data obtained from LVAD-supported individuals. These findings are significant as they provide proof-of-concept evidence for further exploration of heart sound analysis in LVAD-supported individuals to identify cardiac abnormalities and changes in LVAD support.

## KEY WORDS

## INTRODUCTION

In cases of advanced heart failure (HF), a left ventricular assist device (LVAD) may be implanted to assist heart function, rerouting blood from the left ventricle of the heart into the aorta via a mechanical pump. While conferring a survival benefit, LVAD therapy is also characterized by a high complication rate, with hospital readmission rates of 31% within one month [1] and up to 80% within one year of LVAD implantation [2-4].

Common complications include infection, gastrointestinal bleeding, right heart failure, stroke, and formation of blood clots within the LVAD (or thrombosis) [5]. Due to the high complication rate associated with LVAD therapy, the clinical management of LVAD-supported individuals involves routine clinic visits to assess patient health. Despite this, many LVAD complications are not detected until they are at an advanced stage, when patients present with severe or life-threatening symptoms, resulting in prolonged hospital stays and the need to pursue high risk rescue strategies.

Frequent visits to specialized LVAD centers can be costly. Furthermore, without overt evidence for device malfunction, impending complications may not be readily apparent. Remote monitoring of LVAD-supported individuals could allow physicians to identify complications that arise between clinic visits and initiate treatment to mitigate impending complications. Thus, there is a critical need for point-of-care diagnostic tools that can be deployed in patients' homes to facilitate remote surveillance and early identification of LVAD complications.

Running pumps, including the heart and LVAD, generate sounds with characteristic frequency spectra based on pump structure, speed and loading conditions [6-8]. Changes in the distribution of spectral energy can be analyzed to assess proper function [8-10]. Auscultation, performed by listening to physiological sounds with a stethoscope, is an inexpensive and quick method used by physicians to diagnose various cardiovascular problems. For example, cardiologists assess psychoacoustic features, such as loudness, pitch, and tonality, to identify valvular heart disease [11-13]. However, the dominance of the LVAD sounds in the acoustic frequency spectrum limits the use of auscultation in LVAD-supported individuals. Advances in digital stethoscopes have allowed for high-fidelity digital sound recordings, making acoustic analysis a potential non-invasive remote monitoring approach that could be used to identify LVAD complications between clinical visits. Previous studies have shown differences in the acoustic frequency spectra of LVADs with and without thrombi [14-17], demonstrating the potential diagnostic utility of acoustic analysis in LVAD-supported individuals.

While prior acoustic analysis in LVAD-supported individuals has focused on assessing LVAD function, we focus on the analysis of underlying heart sounds that are normally obscured by LVAD sounds. We leverage adaptive filtering and knowledge of LVAD and heart sound characteristics to develop an algorithm to extract heart sounds in the presence of the LVAD sounds. Using data obtained *in vivo* from LVAD-supported individuals, we demonstrate that the filtering of LVAD sounds can isolate heart sounds. After application of a state-of-the art heart sound segmentation algorithm [18] to the denoised signals, we analyze their acoustic signatures with reference to known heart sound characteristics using various time and frequency domain representations with the goal of extracting heart sounds.

## BACKGROUND

Auscultation of heart sounds is performed by listening to sounds through the chest wall overlaying the heart, known as the *precordium,* with a stethoscope; sounds recorded from this region will be referred to as *precordial sounds*. Precordial sounds may contain a mixture of physiological signals, predominantly heart and breath sounds, each of which have distinctive time-frequency characteristics [19]. A recording of heart sounds is known as a *phonocardiogram* (PCG).

In automatic heart sound segmentation, the cardiac cycle is split into four stages: the first heart sound or S1 (closure of atrioventricular valves), systole (heart muscle contraction), the second heart sound or S2 (closure of semi-lunar valves), and diastole (heart muscle relaxation) [11, 18, 20, 21]. Here, S1 and S2 will be referred to as the *fundamental heart sounds* (FHSs), and systole and diastole will be referred to as the *inter-FHS regions*. The FHSs are generally expected to be the loudest sounds occurring in a PCG signal, and exhibit distinct frequency characteristics [11, 21, 22]. The majority of the energy in S1 and S2 is concentrated below 150 Hz, with S2 (24-144 Hz) having a wider frequency range than S1 (24-104 Hz) [22, 23]. It is important to note that the inter-FHS regions may contain extra heart sounds (S3 and S4) and murmurs (due to stenosis, regurgitation, etc.) that may be indicative of additional pathology [11].

In LVAD-supported individuals, precordial sounds contain a mixture of physiological and LVAD sounds. The frequency characteristics of LVAD sounds are determined by the pump's rotational frequency, structure and loading conditions. FHSs, by contrast, are determined primarily by cardiac loading conditions. For example, S2 predominantly reflects aortic valve closure that occurs when the aortic pressure exceeds left ventricular pressure. Because the LVAD lowers left ventricular pressure, many LVAD recipients will have an aortic valve that is mostly closed and have an absent or diminished S2. By contrast, patients with ineffective LVAD unloading have higher left ventricular pressures and may have more aortic valve opening and closing, manifesting as a more prominent S2. Thus, longitudinal changes in intrinsic heart sounds may be useful to monitor for changes in LVAD support and an impending complication of LVAD therapy.

## METHODS

The typical framework for heart sound analysis includes signal pre-processing to increase the signal-to-noise ratio (SNR) of underlying heart sounds, segmentation into cardiac cycle stages, and feature extraction for further analysis [24-26]. In this study, it is assumed that the precordial sounds recorded from the two subjects that participated in the study primarily consist of LVAD and heart sounds. After pre-processing, intrinsic precordial sounds and LVAD sounds are separated prior to cardiac cycle segmentation. Signal processing and analysis were performed in MATLAB© (The MathWorks, Inc.).

### A. Data Description

The data were collected from two patients followed by the Duke Mechanical Circulatory Support Program at Duke University Hospital; this work was approved by the Duke Health Institutional Review Board [27]. Both patients, hereafter referred to as Subject A and Subject B, underwent a medical intervention after an LVAD complication. Pre- and post-intervention recordings of precordial sounds were obtained by a physician from each patient's left upper sternal border using a Thinklabs digital stethoscope (Thinklabs Medical LLC, Centennial, CO). During playback of the recordings, heart sounds, LVAD noise, and to a lesser extent, breath sounds, were observed, with breath sounds being more pronounced in Subject A's recordings.

Each subject's LVAD model, pump operating speed and rotor parameters were noted and used to characterize the acoustic frequency spectrum of their LVAD. Subject A was implanted with a HeartMate II LVAD (HMII; Thoratec Corporation, Inc., Pleasanton, CA) and underwent a surgical pump exchange for pump thrombosis. The HMII is an axial-flow pump with three rotor blades and a pump speed generally ranging from 8,000-12,000 revolutions per minute (rpm) [28]. Subject A's LVAD had an operating speed of 9,200 rpm pre-intervention and 9,400 rpm post-pump exchange. Subject B was implanted with a HeartWare HVAD (HVAD; Medtronic, Inc., Minneapolis, MN) and had a stent inserted after a thrombus had narrowed the LVAD's outflow graft and compromised blood flow from the device. The HVAD is a centrifugal pump with a wide-bladed rotor design with four blood flow channels. Recommended pump speeds are between 2,400-3,200 rpm [29]. Subject B's LVAD had an operating speed of 3,060 pre-intervention and 3,100 rpm post-stenting.

### B. Signal Processing

#### 1) Pre-processing

The precordial sound recordings were collected at a sampling rate of 48,000 Hz. The recordings were first bandpass filtered between 20-500 Hz to restrict the range of frequencies to the expected range of normal and abnormal heart sounds [30]. For computational simplicity and compatibility with the heart sound segmentation algorithm used in this pipeline [18], the recordings were then downsampled to 1,000 Hz.

#### 2) Signal Source Separation

While pre-processing removed high frequency components of the LVAD sounds, additional processing was needed to attenuate frequency components within the bandpass frequency range of 20-500 Hz. The frequency spectrum generated by an LVAD is dependent upon the pump's rotational speed, which defines the *fundamental frequency* (revolutions per minute/60 Hz) of the pump. Frequency peaks can be expected at multiples, or *harmonics*, of the pump's rotational frequency and, in particular, at multiples of the *blade-passing frequency,* which is determined accordingly [31]:

$$f_{bp} = b \times f \qquad (1)$$

where $f_{bp}$ refers to the blade-passing frequency; $b$ refers to the number of blades or blood flow channels in the LVAD rotor; and $f$ refers to the pump's fundamental frequency.

An adaptive filter with a noise cancellation architecture [32] was used to isolate LVAD noise from heart sounds, or the PCG signal. To estimate the noisy signal component ($n$) within a signal mixture ($s + n$), a noise reference signal ($n_{ref}$) that is both correlated with the noise within the mixture and uncorrelated with the signal(s) of interest is required. The noise estimate $\hat{n}$ is substracted from the signal mixture to obtain a denoised signal, $\hat{s} = (s + n) - \hat{n}$. Through a closed-loop process, the filter weights are adapted iteratively to minimise this signal difference. A normalized least-mean-squares (nLMS) adaptive filter was used, with the filter weights updated accordingly [33]:

$$w(t + 1) = w(t) + \frac{\bar{\mu}}{\varepsilon + n_{ref}{}^T(t)n_{ref}(t)}\epsilon(t)n_{ref}(t)$$

where $w$ is the filter weight vector; $t$ is the current time index; $\bar{\mu}$ is the nominal step size; $n_{ref}$ is the noise reference; $\epsilon = (s + n) - \hat{n}$ refers to the difference between the filter output and the mixture of signals; and $\varepsilon$ is a small constant that prevents division by a very small number.

Here, the input signal mixture is the recording of precordial sounds, the noise within the mixture is the LVAD-generated signal, and the output after adaptive filtering is assumed to be the intrinsic precordial sounds, or denoised PCG signal. A noise reference was generated with a mixture of sinusoids corresponding to the frequencies at the LVAD harmonics of interest falling below 500 Hz, as determined by (1). An nLMS filter with a length of 200 samples was implemented using a filter step size $\bar{\mu}$ experimentally tuned by assessing the attenuation of amplitudes at LVAD-specific frequencies in the denoised signal.

### 3) Signal Segmentation

Finally, a state-of-the-art heart sound segmentation algorithm developed by Springer *et al.*, 2017 [18] was employed to segment the denoised PCG signals. The Springer *et al.*, 2017 algorithm is based on a hidden semi-Markov model (HSMM) that incorporates expected stage durations and logistic regression-based emission probabilities to determine the most likely sequence of stages after analyzing input features extracted from the signal. The input features include: the homomorphic envelogram; the Hilbert envelope; the wavelet envelope; and the power spectral density (PSD) envelope, which are calculated at a sampling frequency of 50 Hz [18]. The parameters of the HSMM were trained using a database of PCG signals collected from a non-LVAD population. Since a database of PCG signals obtained from the LVAD population is unavailable, this model was used with the assumption that the learned PCG signal model will translate across populations.

### C. Post-segmentation Analysis

Due to the lack of cardiac cycle labels for the denoised PCG data, the performance of the segmentation algorithm cannot be directly evaluated. Therefore, existing knowledge of heart sound characteristics was used to assess the validity of the cardiac cycle labels estimated by the segmentation algorithm.

### 1) Springer's Segmentation Algorithm Features

The features extracted by Springer's segmentation algorithm were examined to assess the separability of the labelled cardiac cycle stages. For visualization, the features were averaged over each segmented cardiac cycle stage and the dimensionality of the averaged features was reduced to three dimensions using principal component analysis (PCA).

### 2) Cross-correlation

The validity of the segmentation labels was also assessed by comparing similarities of signals within the same stage and between stages. We expect S1 and S2 signals to be more similar to themselves than to any other stage, and to be more similar to each other than to the generally quieter inter-FHS regions. The cross-correlation metric measures the similarity between two signals as a function of the delay of one relative to the other and is calculated accordingly [34]:

$$R_{x,y}[l] = E\{x[t]y^*[t-l]\}$$

where $R_{x,y}[l]$ denotes the cross-correlation between two signals, $x$ and $y$, at a lag $l$ of one signal relative to the other; $t$ is the time index; $*$ denotes the complex conjugate of a signal; and E{} denotes the expected value. The cross-correlation metric accounts for non-alignment and amplitude scale differences (via a normalized measure) between signal pairs. The maximum cross-correlation between two signals $\left(R_{x,y}^{max}\right)$ across times lags was determined for each signal pair. The average maximum within-stage and between-stage cross-correlations were computed based on pairings of signals of the same stage and across two stages, respectively.

### 3) Power Spectral Density Estimation

Frequency representations of the signals were obtained to assess the removal of LVAD harmonics of interest after adaptive filtering and compare the frequency characteristics of the segmented signals with expected heart sound characteristics. A periodogram was used to estimate the power spectral density (PSD) of a signal against frequency accordingly [34]:

$$\hat{P}_x\left(e^{j\omega}\right) = \frac{1}{N}\left|X\left(e^{j\omega}\right)\right|^2$$

where $\omega$ is the angular frequency; $\hat{P}_x\left(e^{j\omega}\right)$ is the power spectral density estimate; $X\left(e^{j\omega}\right)$ is the Fourier transform of the signal; and $N$ is the signal length in samples.

Welch's windowing and averaging method, which reduces the variance of the estimated PSD, was used to compute the PSD estimates [35]. The Hann function was used for windowing: when the entire PCG signal was processed, a window length of 0.2s and window overlap of 50% were used to deliver an appropriate frequency resolution; when individual cardiac cycle stages were processed, a window length of 75% of the signal length, selected due to the small overall signal length, and an overlap of 50%, were used.

### 4) Continuous Wavelet Transform

Time-frequency representations of the denoised PCG signals were obtained to describe changes in the signals' frequency content over time, enabling comparisons between the characteristics of these signals and known characteristics of heart sounds. A one-dimensional (1-D) continuous wavelet transform (CWT) performed with the transform calculated accordingly [36]:

$$\gamma(s,\tau) = \int f(t)\psi_{s,\tau}^*(t)dt$$

where $\gamma(s,\tau)$ is the continuous wavelet transform; $f(t)$ is the original signal that is decomposed into a set of basis functions s; $\psi(t)$ generated from a mother wavelet; $*$ denotes the complex conjugate; and $s$ and $\tau$ denote the scaling and translation of the mother wavelet, respectively. The Morlet wavelet was used [37] due to its similarity to the fundamental heart sounds [38, 39] and its ability to provide a reliable time-frequency representation for PCG signal analysis [40].

## RESULTS

Each subject's precordial sound recordings were processed and segmented using the signal processing pipeline previously described above, with Figure 1 showing the effect of filtering on the post-intervention recordings. The PSD estimates of the raw signals for Subjects A and B, shown in Figure 1c and g, respectively, indicate peaks at the pump harmonics. After adaptive filtering, pump-specific frequencies are notably attenuated, and the majority of the signal power in each denoised post-intervention PCG is below 200 Hz, which is consistent with the expected range of heart sounds. The impact of adaptive filtering is more salient in Subject B's recordings (Figure 1g and h), likely due to the lower LVAD rotational frequency resulting in more noise below 500 Hz.

Figure 1d and h show the magnitude scalograms of the CWT of the denoised PCG signals for Subjects A and B, respectively. In general, there is more energy in the labelled FHS segments than in inter-FHS regions (systole and diastole). In Subject A's signals, there appears to be some high frequency activity in the inter-FHS regions, which may be attributed to breath sounds or regurgitation sounds. Also, labelled S2 segments in Subject A are relatively muted, potentially indicative of intermittent or absent S2, which is consistent with enhanced left ventricular unloading after dysfunctional LVAD exchange based on anecdotal information from clinical experts. Overall, adaptive filtering removed LVAD noise and improved the SNR of the PCG signals with minimal impact on heart sounds.

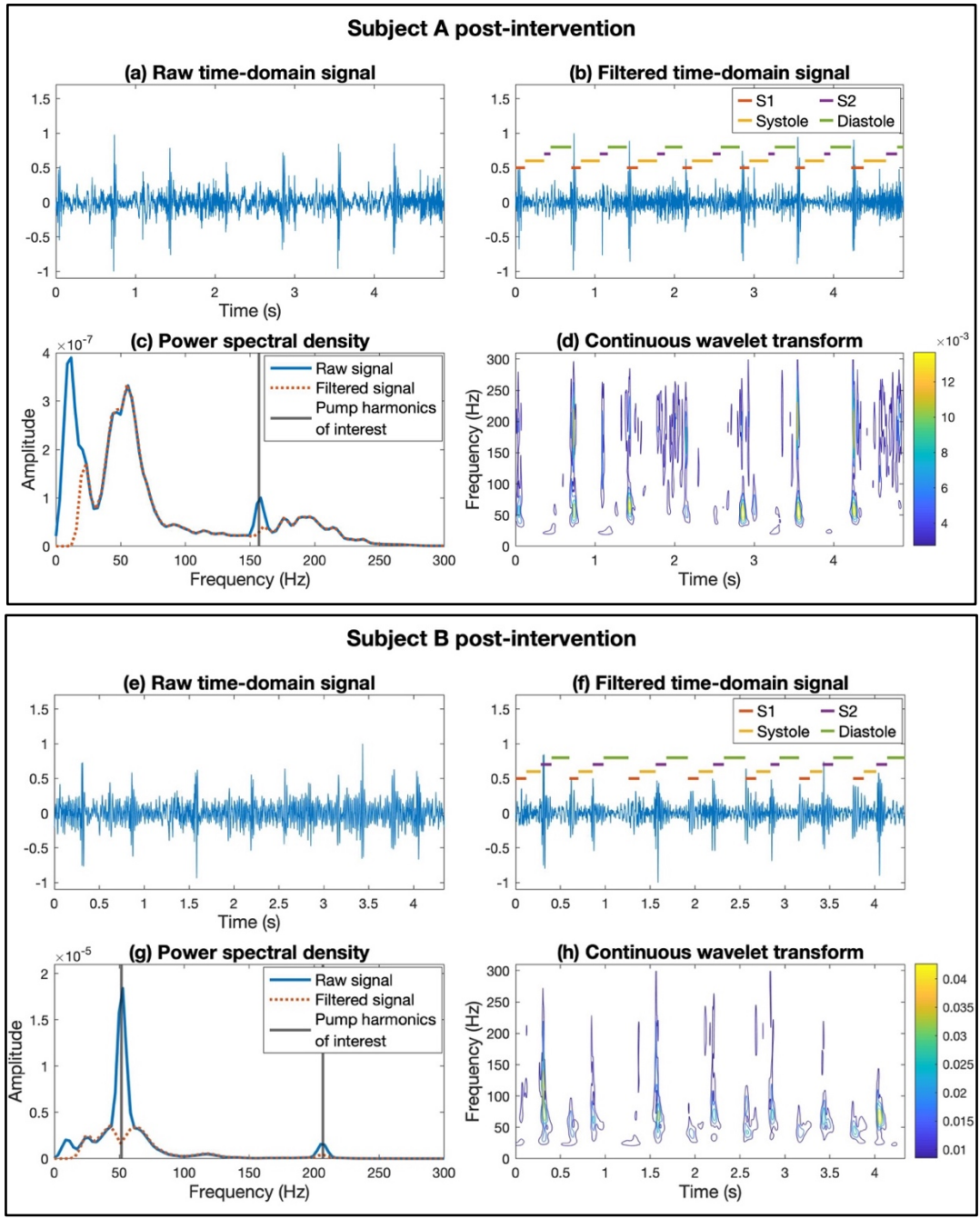

Figure 1. Time- and frequency-domain representations of post-intervention signals collected from Subjects A (top panel) and B (bottom panel). (a, e) Raw time-domain PCG (b, f): Filtered time-domain PCG signal with superimposed labels. (c, g): Power spectral density (PSD) estimates of raw and filtered recording, with the LVAD harmonic frequencies noted; (d, h): Magnitude scalogram contour of the continuous wavelet transform (CWT) of the PCG signal post-adaptive filtering.

The performance of the segmentation algorithm was assessed qualitatively by analyzing temporal and spectral characteristics of the labelled cardiac cycle stages. Figure 2a-b and 2e-f show the PCA visualization of the features extracted by Springer's segmentation algorithm from the denoised PCG signals for Subject A. For Subject A, the features from labelled S1 segments have distinct clusters, while the labelled S2 segments cluster together with the labelled inter-FHS segments (Figure 2a-b). This suggests that Subject A's labelled S2 sounds are similar to labelled inter-FHS segments, in accordance with the previous observation of muted S2 sounds (see Figure 1a-b). In contrast, in Subject B's PCA visualization, labelled FHS segments cluster distinctly from labelled inter-FHS segments (Figure 2e-f). The averaged maximum cross-correlations of the temporal profiles of cardiac cycle stage segments for Subjects A and B are shown in Figure 2c-d and g-h, respectively. For each subject, the highest within-stage correlations are observed for FHS segment pairs (S1-S1 and S2-S2), followed by between-stage correlation of FHS segment pairs (S1-S2), except for cross-correlations of labelled pre-intervention S2 segments for Subject A.

Analyses of PCA clusters and cross-correlation suggest that labelled FHSs are distinct from the inter-FHS regions (except for Subject A's labelled S2 segments) and from each other, and labelled inter-FHS segments generally exhibit lower within- and between-stage correlations. For Subject A's post-intervention signals, while features from labelled S2 segments cluster with inter-FHS segments (Figure 2b), the temporal profiles of S2-S2 segment pairs are noticeably higher than S2-inter-FHS segment pairs (Figure 2d). This finding suggests that alternative signal representations are needed to better characterize and reveal differences between cardiac cycle stage segments.

Knowledge of the expected amplitude and frequency characteristics of the FHS and inter-FHS regions was leveraged to inform our assessment of the validity of the segmentation algorithm. The PSD estimates of the cardiac cycle stage segments for Subject A and B are shown in Figure 3a and b, respectively. Except for the muted S2 sounds in Subject A, labelled inter-FHS segments (systole and diastole) are much lower in amplitude than the FHSs in each subject's recordings, which matches expectations that the FHSs are the dominant sounds in PCGs. The relatively low amplitudes of labelled S2 sounds in Subject A again indicate faint or absent S2 sounds (Figure 3a). For Subject B, three attributes of the labelled FHSs in Figure 3b strongly suggest accurate segmentation. First, the dominant frequencies of labelled S2 sounds are higher than those of labelled S1 sounds [11, 23]. Next, the frequency ranges of the labelled S2 segments are broader than those of the labelled S1 segments [22]. Finally, the amplitudes of labelled S2 sounds are higher than those of labelled S1 sounds, which is generally expected from recordings made at the left upper sternal border [13].

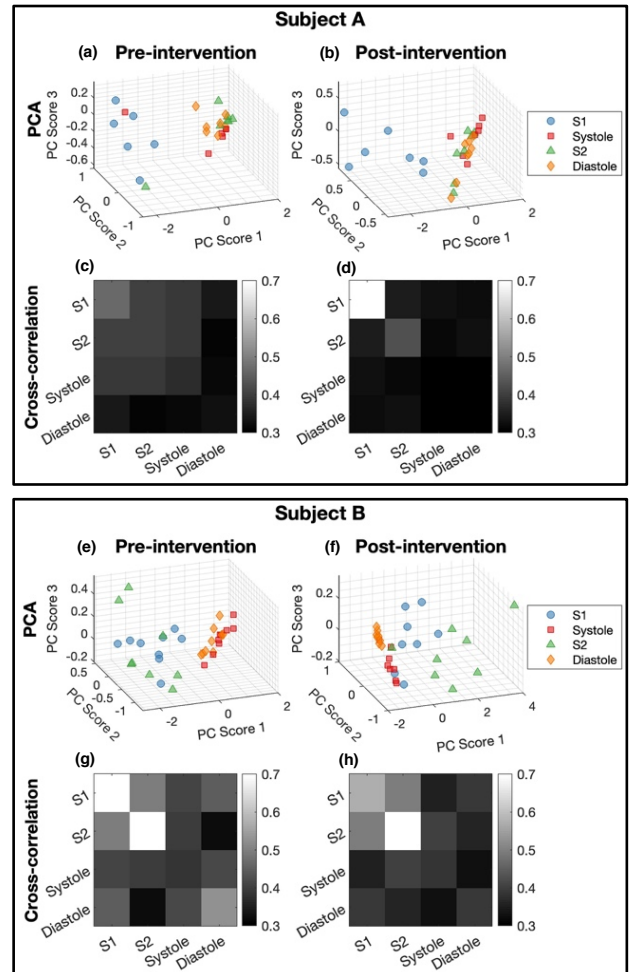

Figure 2. Principal component analysis (PCA) and average maximum cross-correlation representations of cardiac cycle stage segments obtained from pre- and post-intervention PCG signals of Subjects A (left panel) and B (right panel). In each panel, the top row shows the 3D principal component representations of Springer segmentation algorithm features for pre-intervention (a, e) and post-intervention (b, f) signals, labelled by assigned cardiac cycle stage; and the bottom row shows the average maximum cross-correlation values within- and between- cardiac stages for pre-intervention (c, g) and post-intervention (d, h) signals. Higher cross-correlation values indicate greater similarity between stages.

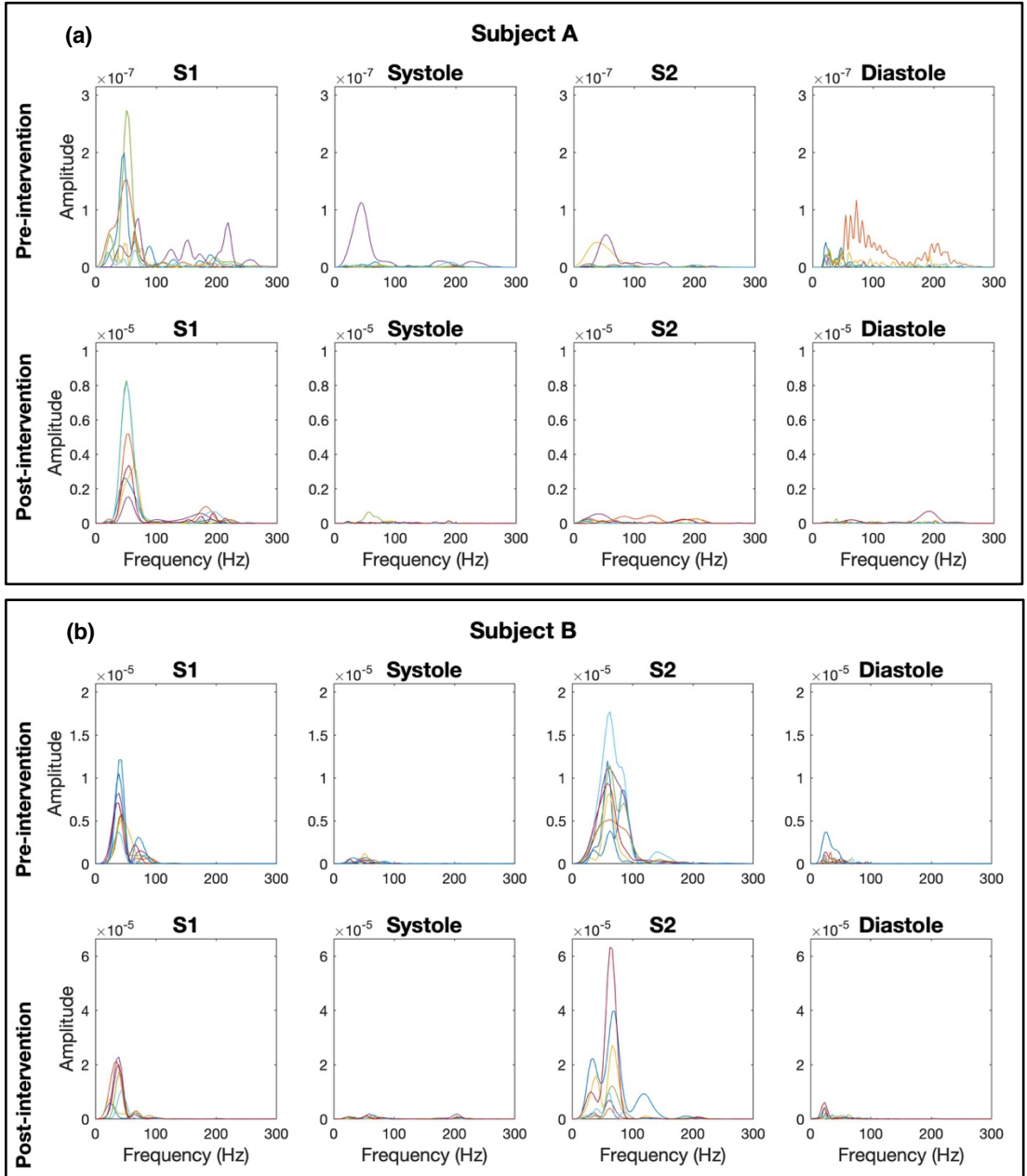

Figure 3. Power spectral density estimates of labelled cardiac cycle stage segments in pre- and post-intervention PCG signals from (a) Subject A and (b) Subject B.

## CONCLUSIONS

This work developed and evaluated a signal processing pipeline to extract and analyze heart sounds in LVAD-supported individuals. Precordial sounds were pre-processed and adaptive filtering applied to attenuate LVAD-specific frequencies and enhance underlying heart sounds. A state-of-the-art segmentation algorithm was used to identify and label the cardiac cycle stages within the denoised PCG signals. While the performance of the segmentation algorithm could not be directly evaluated due to the absence of ground truth labels, various time and frequency analysis techniques were employed to assess the validity of the algorithm's cardiac cycle label outputs. In general, the fundamental heart sounds (S1 and, where present, S2) could be distinguished from the inter-FHS regions (systole and diastole), and the time-frequency characteristics of the FHS segments generally showed agreement with published characteristics.

Acoustic analysis has the potential to be a relatively low-cost diagnostic tool for remote and real-time monitoring of heart function in LVAD-supported individuals. While the work presented here utilized a limited amount of data, this is, to our knowledge, the first preliminary evidence of the feasibility of automated heart sound analysis in LVAD-supported individuals. Our algorithm development and analysis relied on leveraging prior work on heart sound analysis in individuals without LVAD support, with the assumption that heart sound characteristics are similar across individuals with and without LVAD support. However, this assumption might not necessarily provide the best context for analyzing the potential diversity of heart sounds that can be observed in the population of LVAD-supported individuals, as is demonstrated by the between-subject variabilities observed in this study. In addition, the presence of other physiological sounds, e.g., lung sounds and sounds that can be generated due to pump-heart interactions, may impact algorithm performance. Future work includes developing a repository of precordial sounds from individuals with LVADs to develop more robust signal processing algorithms and better characterize acoustic signatures of heart sounds in the population of LVAD-supported individuals.

## ACKNOWLEDGEMENTS

X. Chen received grant support from the Duke Institute for Health Innovation. The authors would like to thank Clive Smith and Thinklabs Medical LLC for providing the digital stethoscopes and for their helpful input.